\documentclass[manuscript,acmsmall,authorversion]{acmart}

\usepackage{balance}
\usepackage[inline]{enumitem}
\usepackage{graphicx}
\usepackage{textcomp}
\usepackage{tikz}
  \usetikzlibrary{shapes,arrows,automata,shadows,positioning,arrows.meta}
\usepackage{multicol}
\usepackage{multirow}
\usepackage{newfloat}
\usepackage{placeins}
\usepackage{relsize}
\usepackage{slashbox}

\usepackage{selnolig}  
\usepackage{soul}
\usepackage{upquote}

\usepackage{xspace}

\usepackage{listings}

\usepackage{tcolorbox}

\newcommand{\squeezeup}{\vspace{-4mm}}

\tcbuselibrary{listings}
\definecolor{c_keyword}{rgb}    {0.0, 0.0, 0.6}
\definecolor{c_macro}{rgb}      {0.5, 0.4, 0.0}
\definecolor{c_stl}{rgb}        {0.0, 0.5, 0.5}
\definecolor{c_remark}{rgb}     {0.0, 0.6, 0.0}
\definecolor{c_cuda}{rgb}       {0.6, 0.1, 0.4}
\definecolor{c_user_macro}{rgb} {0.3, 0.2, 0.1}
\definecolor{c_string}{rgb}     {0.1, 0.6, 0.1}

\definecolor{c_lst_colback}{rgb}{.98, 0.98, .98}

\newcommand{\code}[1]           {\texttt{\small#1}}
\newcommand{\codekeyword}[1]    {\code{\color{c_keyword}#1}}
\newcommand{\codemacro}[1]      {\code{\color{c_macro}#1}}
\newcommand{\codestl}[1]        {\code{\color{c_stl}#1}}
\newcommand{\coderemark}[1]     {\code{\color{c_remark}#1}}
\newcommand{\codecuda}[1]       {\code{\color{c_cuda}#1}}

\newcommand{\codestring}[1]     {\code{\color{c_string}#1}}

\newcommand{\ccode}[1]           {\texttt{#1}}
\newcommand{\ccodekeyword}[1]    {\ccode{\color{c_keyword}#1}}
\newcommand{\ccodemacro}[1]      {\ccode{\color{c_macro}#1}}
\newcommand{\ccodestl}[1]        {\ccode{\color{c_stl}#1}}

\newcommand{\ccodecuda}[1]       {\ccode{\color{c_cuda}#1}}

\newcommand{\lstfont}[1]{\color{#1}\scriptsize\ttfamily}
\newtcblisting[use counter=figure]{tlisting}[3][]{
    before upper={\parindent0em},
    left    = .5em,
    top     = 0em,
    bottom  = 0em,
    label   = #3,
    listing only,
    listing options  = {
        language     = c++,
        basicstyle   = \scriptsize\ttfamily,
        breaklines   = false,
        escapeinside = {<@}{@>},
    },
    colback = c_lst_colback,
    colbacktitle = yellow!50!red!50,
    colframe     = yellow!50!red!50,
    coltitle     = white,
    fonttitle    = \bfseries,
    title        = Listing \thetcbcounter: {#2},
    #1
    }
    
\newtcblisting[auto counter]{slisting}[2][]{
    boxrule = 0pt,
    top     = -0.3em,
    left    = 0em,
    bottom  = -0.3em,
    right   = 0em,
    listing only,
    listing options  = {
        language     = c++,
        basicstyle   = \scriptsize\ttfamily,
        breaklines   = false,
        escapeinside = {<@}{@>},
    },
    colback = c_lst_colback,
    }

\newcommand{\hdc}                   {\code{hdc}\xspace}
\newcommand{\HDC}                   {\code{HDC}\xspace}
\newcommand{\exptrelaxedconstexpr}  {\code{-\relax-expt-\-re\-lax\-ed-\-const\-expr}\xspace}
\newcommand{\exptextendedlambda}    {\code{-\relax-expt-\-ex\-tend\-ed-\-lamb\-da}\xspace}
\newcommand{\extendedlambda}        {\code{-\relax-ex\-tend\-ed-\-lamb\-da}\xspace}
\newcommand{\releaseassert}         {\code{re\-lease\_\-as\-sert}\xspace}
\newcommand{\chdc}                  {\ccode{hdc}\xspace}

\newcommand{\return}        {\codekeyword{re\-turn}\xspace}
\newcommand{\constexpr}     {\codekeyword{const\-expr}\xspace}
\newcommand{\requires}      {\codekeyword{re\-quires}\xspace}

\newcommand{\main}          {\codekeyword{main}\xspace}
\newcommand{\cexplicit}     {\ccodekeyword{ex\-pli\-cit}\xspace}
\newcommand{\cnoexcept}     {\ccodekeyword{no\-ex\-cept}\xspace}
\newcommand{\cifconstexpr}  {\ccodekeyword{if const\-expr}\xspace}

\newcommand{\crequires} {\ccodekeyword{re\-quires}\xspace}

\newcommand{\VAARGS}    {\codemacro{\_\_VA\_ARGS\_\_}\xspace}

\newcommand{\macroifdef}{\codemacro{\#ifdef}\xspace}

\newcommand{\DEPAREN}   {\codemacro{DE\-PAREN}\xspace}
\newcommand{\REQUIRES}  {\codemacro{RE\-QUIRES}\xspace}
\newcommand{\has}       {\codemacro{has}\xspace}
\newcommand{\hasxxx}    {\codemacro{has\_xxx}\xspace}
\newcommand{\HASMAKE}   {\codemacro{HAS\_\-MAKE}\xspace}

\newcommand{\cVAARGS}   {\ccodemacro{\_\_VA\_ARGS\_\_}\xspace}
\newcommand{\cREQUIRES} {\ccodemacro{RE\-QUIRES}\xspace}

\newcommand{\stdenableif}   {\codestl{std::enable\_if}\xspace}

\newcommand{\stdabort}      {\codestl{std::abort}\xspace}
\newcommand{\cstdmove}      {\ccodestl{std::move}\xspace}
\newcommand{\cstdforward}   {\ccodestl{std::forward}\xspace}
\newcommand{\cstdfunction}  {\ccodestl{std::function}\xspace}

\newcommand{\hdwarningdisable}      {\codecuda{\#hd\_\-war\-ning\_\-dis\-able}\xspace}
\newcommand{\nvexeccheckdisable}    {\codecuda{\#nv\_\-exec\_\-check\_\-dis\-able}\xspace}
\newcommand{\host}                  {\codecuda{\_\_ho\-st\_\_}\xspace}
\newcommand{\device}                {\codecuda{\_\_de\-vi\-ce\_\_}\xspace}

\newcommand{\cudaglobal}            {\codecuda{\_\_glo\-bal\_\_}\xspace}
\newcommand{\cudadevicesynchronize} {\codecuda{cudaDeviceSynchronize}\xspace}
\newcommand{\trap}                  {\codecuda{\_\_trap}\xspace}
\newcommand{\chost}                 {\ccodecuda{\_\_ho\-st\_\_}\xspace}
\newcommand{\cdevice}               {\ccodecuda{\_\_de\-vi\-ce\_\_}\xspace}
\newcommand{\cconstant}             {\ccodecuda{\_\_con\-stant\_\_}\xspace}
\newcommand{\cshared}               {\ccodecuda{\_\_shared\_\_}\xspace}
\newcommand{\ctrap}                 {\ccodecuda{\_\_trap}\xspace}

\newcommand{\CUDACC}                {\codecuda{\bfseries{\_\_CU\-DA\-CC\_\_}}\xspace}
\newcommand{\CUDAARCH}              {\codecuda{\bfseries{\_\_CU\-DA\_\-AR\-CH\_\_}}\xspace}
\newcommand{\CUDACCRELAXEDCONSTEXPR}{\codecuda{\bfseries{\_\_CUDA\-CC\_\-RE\-LAXED\_\-CONST\-EX\-PR\_\_}}\xspace}
\newcommand{\cCUDAARCH}             {\ccodecuda{\bfseries{\_\_CU\-DA\_\-AR\-CH\_\_}}\xspace}

\newcommand{\nvcc}                  {\emph{nvcc}\xspace}
\newcommand{\clang}                 {\emph{clang}\xspace}
\newcommand{\gcc}                   {\emph{gcc}\xspace}
\newcommand{\UB}                    {\emph{UB}\xspace}

\makeatletter
\@ifpackageloaded{hyperref}{}{\usepackage{hyperref}}
\makeatother
\newcommand{\compilermsg}[1]{{\footnotesize\emph{``\texttt{#1}}}''}

\usepackage{tikz}

\newcommand{\XX}[1]{{\color{red} \noindent\framebox{\footnotesize\textbf{XX}}{ \textrm{\textbf{~#1~}}}}}

\newdimen\hfuzz  

\begin{document}

\title[\chost \cdevice]{\chost \cdevice{} - Generic programming in Cuda}

\author{Thomas Mejstrik}


\begin{abstract}
We present patterns for Cuda/C++ to write save generic code which works both on the host and device side.
Writing templated functions in Cuda/C++ both for the CPU and the GPU
bears the problem that in general both \host{} and \device functions are instantiated,
which leads to lots of compiler warnings or errors.
\end{abstract}

\maketitle

\section{Introduction}

\subsection{Motivation}
When writing generic code in Cuda for both the device side (GPU),
as well the host side (CPU), one faces easily a problem:
Some generic code only works on one of the two sides.
Yet, the Cuda language has no means of specifying for which target something shall be instantiated.
This leads in the best case to 
$(+)$~compiler errors,
$(\circ$)~pages of compiler warnings (which then hide the important warnings),
$(-)$~a program which compiles but crashes when run,
or in the worst case
$(--)$~a program which does the wrong thing ($\rightarrow$ \emph{Undefined Behaviour}).

This paper describes patterns to solve the problem for 
\emph{templated functions and member functions}.
We shortly discuss lambdas, but since Cuda has no good support for templated lambdas yet,
they are of not much use to solve our problem.
We do not discuss how constants can be used both in host and device code efficiently.

\subsection{Background and notation}
\subsubsection{Cuda background}
We summarize some Cuda and C++ facts, which are important for this paper.

\paragraph{Function execution space specifiers}
\label{sec_function_exectuion_space_specifiers}
Cuda distinguishes between \host, \device, and \cudaglobal functions~\cite{cuda_doc_execution_space}:
\host functions run on the CPU,
\device and \cudaglobal functions run on the GPU.
Unsurprisingly,
\host functions are allowed to call \host and \cudaglobal functions,
\cudaglobal and \device functions are allowed to call \cudaglobal and \device functions.
Calls which are not allowed are called \emph{stray} calls in the following.

Functions without any \host nor \device annotations (we will call those \emph{undecorated}),
are implicitly considered as \host functions.

We will refer to code in \host functions as \emph{host code},
and to code in \device and \cudaglobal functions as \emph{device code}.
We will furthermore do not discuss \cudaglobal functions any more,
since for this papers goal they behave the same as \device functions.

Unfortunately, nvcc 12 is quite inconsistent in compiling stray function calls,
see Listing~\ref{lst_problem_t}.
The first (commented out) call of \code{D{}.call()} leads correctly to a compilation error.
If the call is done via a wrapper with the (commented out) code of \code{func< D >()} nvcc 12 issues even no warning.
The third call \code{func< H >()} gives a compilation warning, although the call is totally valid\footnote{%
One could argue that producing a warning here is sensible.
}.

From our test we can at least infer that nvcc behaves correctly when no \host \device functions are in play,
i.e.\ when a function is only \host \emph{or} \device decorated, see Listing~\ref{lst_host_or_device}.


\paragraph{Kernels}
\cudaglobal functions, also also called \emph{kernels},
are the entry point for \host functions
to start something on the GPU.
\cudaglobal functions are called using the \emph{triple chevron} 
execution configuration syntax \codecuda{<{}<{}< ... >{}>{}>},
which specifies (among other things) the number of threads used~\cite{cuda_doc_kernel}.
In Listing~\ref{lst_kernel}, a \cudaglobal function on $4 \times 3$ threads is started,
and each kernel calls 2 times the \code{\device print()}  function, which prints a dot.
Thus, in total 24 dots are printed.

\paragraph{Cuda error checking}
Since Cuda has no support for exceptions in device code,
error handling is done via checking error codes.
Furthermore, an error on the GPU usually never leads to a termination of the program,
just to undefined behaviour if the GPU is used subsequently.
Thus, after each (potentially failing) Cuda call, one should to check the error code.
We will use \cudadevicesynchronize for this task.\footnote{%
In production code error checking should be done differently.}
A return code of \code{0} means success.

\paragraph{nvcc}
Nvidia Cuda compiler \nvcc processes Cuda code in two steps.
First, it processes all device specific stuff by itself, and then passes
a processed version of the code to the host compiler, usually \gcc or \clang.

\paragraph{Lambdas}
A C++ lambda (anonymous function) is started with the capture clause \code{[/*vari\-ables*/]},
followed by the list of arguments in parentheses \code{(/*args*/)},
the function body \code{\{/*body*/\}}, and
the return type \code{-> /*rettype*/}.
The return type of a lambda can be omitted, if the compiler can deduce it.
Thus, a lambda without any code is \code{[]()\{\}}.\footnote{%
It is even allowed to omit the argument list in a lambda and write \code{[]\{\}},
although we will not use this ``feature''.}

Undecorated lambdas in Cuda code inherit the function execution space of its enclosing scope.
E.g., is the lambda defined in a \device function, then the lambda is for device code.

\paragraph{Extended Lambdas}
These are lambdas which are decorated. If one wants to use them, one has to compile with the flag
\exptextendedlambda or \extendedlambda.

Although their use is not experimental any more, they should be considered with caution.
Extended lambdas are wrapped in a \cstdfunction~\cite{cuda_doc_ext_lambda},
and thus may not be able to get inlined,
in particular their performance may be worse than that of a traditional lambda.
Furthermore, they have lots of restrictions.
In particular, it is not possible to write generic extended lambdas.


\paragraph{Line continuation}
A backslash as last character on a line, is a line continuation.
Since macros must be written in exactly one line,
using \code{`\textbackslash`}
this is a way to split out long macros into more lines.

\paragraph{Omitting return in main}
It is allowed by the standard to omit the \return statement in the \main function.

\paragraph{Undefined Behaviour (\UB)}
C++ has a ``feature'' called \UB, meaning \emph{undefined behaviour}.
Whenever an operation results in \UB, the behaviour of the program 
(starting from that particular operation) is not defined any more.
In other words: Anything can happen.
For an example of UB see Listing~\ref{lst_problem_t}.

The compiler is not mandated to, nor it is able to,
diagnose all potential appearances of \UB.
On the other hand, just because something is \UB,
the compiler is not mandated to generate broken code.
Most cases of \UB are described in the C++ standard.

\subsubsection{Miscellanea}
\paragraph{Condensed code samples}
In order to save space in the listings,
we use a style not recommended for production code.
In particular we use (very) short identifiers,
omit the inclusion of headers (e.g.~\codestring{<cas\-sert>} )
or headers of functions described in this paper (e.g.\ \codestring{"re\-lease\_as\-sert.h"}).
Types only compatible on the host are usually named \code{`H`},
types only for the device with \code{`D`}.

\paragraph{Targeted systems}
We target the C++ versions 14, 17, and 20,
as well as Cuda versions later and including Cuda 9.\footnote{%
All examples are thouroughly tested with
\gcc 11.3, \clang 14.0 and \nvcc 12.1.
For older compiler versions \emph{godbolt}~\cite{godbolt} was used.
We did not try to compile the Cuda code with \clang alone.}

We target multiple C++ versions
because often (Cuda) developers do not have access to the latest C++ version,
the Cuda standard lacks behind the C++ standard usually for years
and/or (Cuda) compilers may have bugs.

\twocolumn

\begin{figure}
\begin{tlisting}{Behaviour for \host or \device functions}{lst_host_or_device}
struct H { __host__ void call() {} };
struct D { __device__ void call() {} };

__device__ void dev() {
  D{}.call();  // OK
//H{}.call();  // error
}

void hst() {
//D{}.call();  // error
  H{}.call();  // OK
}
\end{tlisting}
\Description{}
\squeezeup
\end{figure}

\begin{figure}
\begin{tlisting}{Example kernel}{lst_kernel}
#include <cstdio>

__device__ void print() {
  printf( "." );
}

__global__ void kernel( int N ) {
  for( int n = 0; n < N; ++n ) {
    print();
  }
}

int main() {
  kernel<<< 4, 3 >>>( 2 );
  return cudaDeviceSynchronize();
}
\end{tlisting}
\Description{}
\squeezeup
\end{figure}

\section{The patterns}

\subsection{Running example}
Our running example is presented in Listing~\ref{lst_problem_t}.
There, a function is templatized for some user defined type (e.g.\ a matrix class).
In \main, the function \code{func} is called with the type \code{H}.
Thus (according to the Cuda specification)
both the \host and the \device version of \code{func} is instantiated with the type \code{H}.
But, \code{H}'s function \code{call} is only a \host function,
and thus, when compiled with \nvcc we get the warning:
\compilermsg{%
calling a \_\_host\_\_ function ("H::\-c\-a\-l\-l()")
from a \_\_host\_\_ \_\_device\_\_ function ("func< ::H> ")
is not allowed}.

Note though, the program is well formed and does what is expected --
it returns \code{3}.
Things are different, when we would instead call \code{func<~D~>()}
(the line marked with \coderemark{//UB}),
which directly leads to undefined behaviour.
On the test system, the program then returns \code{1} (instead of \code{2}).
Interestingly, \nvcc (versions 9 to 12) does not emit any compiler warning in this case.\footnote{%
A bug report was sent to Nvidia concerning this example.
One can hope that this gets fixed in a future \nvcc version.
}

Compiling this example directly with \gcc or \clang will fail,
due to the presence of Cuda specific keywords and functions,
of which the host compiler cannot make use.

\subsubsection{Goal}
In view of the problems described so far we would like to have
\begin{itemize}
\item a compilation error whenever there is a stray function call
\item no compilation warnings due to wrongly instantiated function templates.
\end{itemize} 

\begin{figure}
\begin{tlisting}{Problem T}{lst_problem_t}
struct H {
  __host__ int call() { return 3; }
};

struct D {
  __device__ int call() { return 2; }
};

template< typename T >
__host__ __device__
int wrap() {
  return T{}.call();
}

int main() {
//return D{}.call();  // error
//return wrap< D >(); // no warning, UB
  return wrap< H >(); // warning
}
\end{tlisting}
\Description{}
\squeezeup
\end{figure}

\subsection[User defined host device macros]{User defined \host \device macros\footnote{%
This pattern alone does not solve our problem,
and usually needs to be employed with Solution~\ref{sec_hst_dev_everything}
(the next one in line).
They are split up in two, since they do not belong together logically.
}}
\label{sec_cudatags}

\subsubsection{Context}
Code shall be useable with Cuda and non-Cuda compilers,
and the code does not contain any Cuda specific language constructs.

\subsubsection{Solution}
As already discussed, Cuda code needs to be annotated with
\host, \device, etc..
But, if one wants to write a program which shall also be able to get compiled by non-Cuda compilers,
these keywords are not defined, and thus compilation fails.

There are two nearby solutions for this task:

\paragraph{Solution A}
One defines macros which expand to \host, \device, etc... when a Cuda compiler is used,
and expand to nothing for non Cuda compilers.
See Listing~\ref{lst_cudatags_1}.

This solution has two disadvantages:
Other developers need to know which macros to use, and
the global namespace gets clobbered with names.

\paragraph{Solution B}
The second solution seems much more beautiful, see Listing~\ref{lst_cudatags_2}.
The big plus of it is, 
that it does not require any additional knowledge by other developers.

If the code is not compiled with \nvcc, then one defines macros \host, \device
which expand to nothing.
When the code is compiled with \nvcc no additional macros are defined.
To decide whether the Cuda specific macros need to be defined one can check
the \CUDACC macro -- it is only defined when \nvcc steers the 
compilation process of Cuda source files.
In particular, the macro is still defined 
after \nvcc passes control over to the host compiler
(which than compiles the preprocessed source files from which
\nvcc stripped away all Cuda specific language constructs).

Note though, Solution~B  (strictly speaking) has \UB since identifiers starting
with two underscores are reserved to be used by the compiler~\cite[global.names]{cppdraft}.
Yet, it is very unlikely that Solution~B will break any code.
Indeed, if a compiler shall be a compatible host compiler for \nvcc,
then the identifiers \chost, \cdevice, etc... must not be used by the host compiler.
Thus, a potential host compiler will not use these identifiers for its own purposes.
We suggest, for large library code bases to use Solution~A, otherwise use Solution~B.
Furthermore, legacy code has lots of identifiers starting with two underscores.
So, a compiler will do nothing bad to code with such identifiers whenever possible.

\begin{figure}
\begin{tlisting}{Ugly, but safe, \host \device macros}{lst_cudatags_1}
#ifndef CUDATAGS_A
  #define CUDATAGS_A
  #ifndef __CUDACC__
    #define CUDA_HOST
    #define CUDA_DEVICE
  #<@\texttt{\color{c_macro}else}@>
    #define CUDA_HOST   __host__
    #define CUDA_DEVICE __device__
  #endif
#endif

// Example: function definition
CUDA_HOST CUDA_DEVICE void func2() {}
\end{tlisting}
\Description{}
\squeezeup
\end{figure}

\begin{figure}
\begin{tlisting}{Easy, but \UB, \host \device macros}{lst_cudatags_2}
// cudatags.h
#ifndef CUDATAGS_B
  #define CUDATAGS_B
  #ifndef __CUDACC__
    #define __host__
    #define __device__
  #endif
#endif

// Example: function definition
__host__ __device__ void func1() {}
\end{tlisting}
\Description{}
\squeezeup
\end{figure}

\subsubsection{Assessment}
\begin{itemize}
\item[$+$] Easy to use.
\item[$-$] Only works, if no Cuda specific features are used.
\end{itemize}

\subsection{\host \device everything}
\label{sec_hst_dev_everything}

\subsubsection{Context}
Code shall be useable with Cuda and non Cuda compilers,
and the code does not contain any Cuda specific features.

\subsubsection{Solution}
We add \host \device annotations to all functions.

The code in Listing~\ref{lst_solution_host_device_everything}
compiles both in Cuda and non-Cuda compilers without warning.

\begin{figure}
\begin{tlisting}{Solution: \host \device}{lst_solution_host_device_everything}
#include "cudatags_b.h"

struct S {
  __host__ __device__
  static void value() {}
};

template< typename T > 
__host__ __device__
void func() {
  T::value();
}

int main() {
    func< S >(); 
}
\end{tlisting}
\Description{}
\squeezeup
\end{figure}

\subsubsection{Assessment}
\begin{itemize}
\item[$+$] Solution~A: Easy to use, straight forward, not bug prone and has \UB.
\item[$\circ$] Solution~B: No \UB.
\item[$\circ$] A list of Cuda keywords has to be maintained manually. 
This list is short in general
for code which makes sense both in host and device code.
\item[$-$] Not always possible. The host and device implementation may be different or
there may not be any implementation for either host or device code.
\end{itemize}

\subsubsection{Known Usages}
\begin{itemize}
\item Boost/Utility~\cite{boost_utility} uses a combination\\[-2ex]
\begin{slisting}{}
#define BOOST_GPU_ENABLED \
    __host__ __device__
\end{slisting}

\item Eigen~\cite{eigen} uses a variant\\[-2ex]
\begin{slisting}{}
#ifdef /* some Eigen macro */
  #define EIGEN_DEVICE_FUNC \
    __host__
#<@\texttt{\color{c_macro}else}@>
  #define EIGEN_DEVICE_FUNC \
    __host__ __device__
#endif
\end{slisting}

\end{itemize}

\subsection{\macroifdef blocks with \CUDAARCH}
\subsubsection{Context}
Code shall be useable with Cuda and non-Cuda compilers,
but the necessary implementations differ.
\subsubsection{Solution}
Use preprocessor macros to determine whether we are in device or host code,
see Listing~\ref{lst_solution_release_assert}.

In order to make different function bodies for host and device code, we use the macro \CUDAARCH.
\CUDAARCH is defined when device code is compiled,
i.e.\ after \nvcc passes control over to the host compiler \CUDAARCH is not defined.

\paragraph{release\_assert()}
Listing~\ref{lst_solution_release_assert} presents the function \releaseassert
which we will use subsequently.
\releaseassert takes an argument of type \codekeyword{bool}.
If the argument is \codekeyword{false} then a runtime error is triggered.
In device code this is done by calling \trap, which aborts the kernel execution.\footnote{%
The error code produced by calling \ctrap seems to be (at least) compiler dependent.
In our tests, 
Cuda 9 gave error number 4 (\emph{cudaErrorCudartUnloading}),
Cuda 10 to Cuda 12 gave error number 207 (\emph{cudaErrorArrayIsMapped}).},
in host code by calling \stdabort.
Note that, in production code one may want to use less sever means,
and/or want to print out debug info to \code{stderr}
beforehand~\cite{cuda_assert,cuda_errorcheck}.

\begin{figure}
\begin{tlisting}{Solution: \macroifdef block with \CUDAARCH: \releaseassert}{lst_solution_release_assert}
#include <cstdlib>

#ifdef __CUDA_ARCH__
static constexpr bool cuda_arch = true;
#<@\texttt{\color{c_macro}else}@>
static constexpr bool cuda_arch = false;
#endif

__host__ __device__
void release_assert( bool flag ) {
  if( !flag ) {
    #ifdef __CUDA_ARCH__
    __trap();
    #<@\texttt{\color{c_macro}else}@>
    std::abort();
    #endif
  }
}
\end{tlisting}
\Description{}
\squeezeup
\end{figure}

\subsubsection{Assessment}

\begin{itemize}
\item[$+$] Easy to understand and use
\item[$\circ$] Source code is cluttered with preprocessor directives. 
which is usually not a problem when the function body is short.
\end{itemize}

\subsubsection{Known usages}
\begin{itemize}
\item Eigen
\item Most likely in every larger Cuda codebase
\end{itemize}

\subsubsection{Variant: \macroifdef blocks with \CUDAARCH with no sensible implementation available}
It may happen, that there is no sensible implementation for either host or device code.
In this case, we may abort the program whenever we end up in the wrong path,
see Listing~\ref{lst_solution_ifdef_cudaarch}.\footnote{%
The variable \ccode{cuda\_arch} is usable in device code,
although not being marked with \cconstant or \cshared,
due to~\cite{cuda_doc_const}.}
In that listing, \code{S::value()} is (wrongly) called in device code,
and thus a runtime error on the GPU is triggered.
The runtime error is not reported back to host side,
until one checks the Cuda error code in host code,\footnote{%
This is not entirely true:
\ctrap raises an interrupt on the side which in theory could be handled.}
here again using \cudadevicesynchronize.

One may be tempted to use \code{\codemacro{assert}(\codekeyword{false})} for stray calls.
That would be a bad idea, since it hides the error in production code.
Furthermore, asserts are used in C++ in places where one does 
not want to (or can) pay a performance penalty in production code.
But, if we do a stray function call, we do not have performance problems
-- we have a logical programming error.

\begin{figure}
\begin{tlisting}{Solution: \macroifdef block with \CUDAARCH}{lst_solution_ifdef_cudaarch}
struct S {
  __host__ __device__
  static void value() {
    release_assert( !cuda_arch );
    /* body */
  }
};

template< typename T >
__global__ 
void kernel( T t ) { t.value(); }

int main() {
  kernel<<< 1, 1 >>>( S{} );
  return cudaDeviceSynchronize();
}
\end{tlisting}
\Description{}
\squeezeup
\end{figure}

\subsubsection{Assessment}
\begin{itemize}
\item[$+$] Always usable. Sometimes this solution is the last resort.
\item[$+$] 
No need for preprocessor directives.
\item[$-$] Check of execution space is done at runtime.
\end{itemize}

\subsubsection{Known usages}
\begin{itemize}
\item Eigen
\item Most likely in every larger codebase
\end{itemize}

\subsubsection{\color{red}No solution}

There are some restrictions how \CUDAARCH can be used.
To us the most important are:
The signature of functions, function templates and instantiated function templates,
as well as the arguments used to instantiate function templates
must not depend on whether \CUDAARCH is defined
or not~\cite{cuda_doc_preprocessor_symbols}.

This implies that the solution in Listing~\ref{lst_no_solution_ifdef_cudaarch}
is not a valid one,
since \code{func< S >} is only instantiated when \CUDAARCH is undefined.%
\footnote{%
Although the program in Listing~\ref{lst_no_solution_ifdef_cudaarch} is not a valid one,
nvcc~12.1 gives no warning when compiling it.%
}
$^,$%
\footnote{%
A similar, also invalid idea, is to use a constant dependent on \cCUDAARCH
as a non-type template parameter to some function.
Whether using a \cifconstexpr clause with a constant dependent on \cCUDAARCH 
in the body is unclear. We have no authoritative answer from Nvidia yet.
}

\begin{figure}
\begin{tlisting}{{\color{red}No solution:} \macroifdef block with \CUDAARCH}{lst_no_solution_ifdef_cudaarch}
#include "cudatags.h"

struct S {
  __host__
  static void value() {}
};

template< typename T > 
__host__ __device__
void func() {
  T::value();
}

int main() {
    #ifndef __CUDA_ARCH__  // UB
    func< S >(); 
    #endif
}
\end{tlisting}
\Description{}
\squeezeup
\end{figure}


\subsection{Pragmas \hdwarningdisable and \nvexeccheckdisable}
\label{sol_pragma}

\subsubsection{Context}
One ``knows'' that a certain code path is not possible,
and thus just wants to disable compiler warnings.

\subsubsection{Solution}
The pragmas \hdwarningdisable and \nvexeccheckdisable can be used.

\paragraph{Pragma explanation}
If one of the pragmas is placed in front of a function,
then the compiler does not spill out warnings for bad function calls 
inside that function\footnote{%
The pragmas only affect the function they are placed in front of,
in particular there is no need for a pragma \ccode{hd\_warning\_enable}).
The pragmas also does not affect function calls in subfunctions.
}
(In other words: The caller function must be decorated, not the callee).

But, care should be taken:
These pragmas are undocumented, although Nvidia uses them itself open sourced code.
Their exact behaviour is not clear. In some cases one pragma works,
in some cases the other,
in some cases none.
Even worse, it is reported that a wrong usage of the pragmas
may lead to wrongly compiled code~\cite{cuda_pragma_wrong_code}.
The pragmas could also be changed or removed any time. Indeed,
\hdwarningdisable is an older pragma and seems to get replaced now by \nvexeccheckdisable.
Thus, Code targeted for older compilers may only understand \hdwarningdisable,
whereas future compilers may not understand \hdwarningdisable any more.

\begin{figure}
\begin{tlisting}{Solution: \codemacro{\#pragma}}{lst_solution_pragmas}
struct S {
  static void value() {}
};

#pragma hd_warning_disable
template< typename T > 
__host__ __device__
void func() { T::value(); }

int main() {
  func< S >();
}
\end{tlisting}
\Description{}
\squeezeup
\end{figure}

\subsubsection{Assessment}
\begin{itemize}
\item[$+$] Easy to use
\item[$\circ$] Each function has to be annotated manually.
\item[$-$] These pragmas are undocumented
\item[$-$] Wrong usage of these pragmas may lead to wrongly compiled code~\cite{cuda_pragma_wrong_code}
\item[$-$] May hide programming errors.
A combination of using these pragmas with \releaseassert is thus recommended.
\item[$-$] Even when one ``knows'' that the pragmas can be used at the time when the code is written,
things may change in the future.
\end{itemize}

\subsubsection{Known Usages}
\begin{itemize}
\item Thrust~\cite{thrust}
\item Eigen
\end{itemize}


\subsection{Experimental relaxed constexpr}
\label{sol_exp_relaxed_constexpr}

\subsubsection{Context}
One has a function which is, or can be made, \constexpr
and one compiles with \nvcc.

\subsubsection{Solution}
We decorate the 
function with \constexpr
and compile with the option \exptrelaxedconstexpr~\cite{cuda_doc_exprelaxesconstexpr}.
This allows device code to invoke \host \constexpr functions,
and host code to invoke \device \constexpr functions.
Beware that, this is an experimental feature and only works with \nvcc,
see Listing~\ref{lst_solution_exp_constexpr}.

We assert whether the source code is compiled with  \exptrelaxedconstexpr
by checking whether the macro \CUDACCRELAXEDCONSTEXPR is defined,
and produce a compilation error when not.
This way, the user is informed how to correctly compile the program,
when she attempts to compile it wrongly.

\begin{figure}
\begin{tlisting}{Solution: Experimental relaxed \constexpr}{lst_solution_exp_constexpr}
#ifndef __CUDACC_RELAXED_CONSTEXPR__
#error "Must be compiled with:" \
  "--expt-relaxed-constexpr"
#endif

struct S {
  constexpr static int value() {
    return 42;
  }
};

template< typename T >
__global__
void kernel( T t ) {
    printf( "%i", t.value() );
}

int main() {
  kernel<<< 1, 1 >>>( S{} );
  return cudaDeviceSynchronize();
}
\end{tlisting}
\Description{}
\squeezeup
\end{figure}

\subsubsection{Assessment}~
\begin{itemize}
\item[$++$] Is also applicable to third party \constexpr function, e.g.\ functions in the C++ standard library
\item[$+$] Easy to use.
\item[$+$] Needs minimal changes to the source code.
\item[$-$] Only applicable to \constexpr functions.
\item[$-$] Is an experimental feature:
It is experimental since at least Cuda 8.0 from 2016.
The behaviour of this option may change in future Cuda releases.
\item[$-$] The source code is not self contained any more, 
but needs to be compiled with certain compiler flags.
\item[$-$] It is unclear, whether this feature is compatible with future C++ versions.\footnote{%
Currently, 
Each new C++ standard softens the restrictions to \constexpr functions
E.g., C++20 allows memory allocations in \constexpr functions.}
\item[$-$] Only works with \nvcc.
\end{itemize}

\subsubsection{Known Usages}
\begin{itemize}
\item LBANN~\cite{LBANN}
uses a defensive strategy: 
Is the source compiled with \exptrelaxedconstexpr
they annotate functions with \constexpr, otherwise they annotate them with
\host \device.
\end{itemize}

\subsubsection{Variant with lambdas: C++17}
This solution can also be applied to lambdas (starting from C++17),
see Listing~\ref{lst_solution_exp_constexpr_la}.

There is a notable peculiarity in this Listing, the keyword \constexpr is not present.
This is because starting from C++17,
lambdas are implicit \constexpr whenever they 
happen to satisfy all \constexpr function requirements~\cite[ex\-pr.pr\-im.lam\-bda]{cppdraft}.
This solution does not work with C++14, because C++14 does not allow \constexpr lambdas.

\begin{figure}
\begin{tlisting}{Solution: Experimental relaxed \constexpr for lambdas (C++17)}{lst_solution_exp_constexpr_la}
#ifndef __CUDACC_RELAXED_CONSTEXPR__
#error "Must be compiled with:" \
  "--expt-relaxed-constexpr"
#endif

template< typename Lambda >
__host__ __device__
void func( Lambda la ) {
  printf( "%i", la() );
}

__host__ void wrapper() {
  auto la = [](){ return 3; };
  func( la );
}

__global__ void kernel() {
  auto la = [](){ return 4; };
  func( la );
}

int main() {
  wrapper();
  kernel<<< 1, 1 >>>();
  return cudaDeviceSynchronize();
}
\end{tlisting}
\Description{}
\squeezeup
\end{figure}

\subsubsection{Bad Variant with lambdas: C++14}
A way, without using \constexpr lambdas, 
is to make the lower-level function \constexpr,
see Listing~\ref{lst_solution_exp_constexpr_la_14}.

The example may work or not;
The following behaviour is reported~\cite{problem_lambda14} (and partly double checked):
It works with \nvcc 10.1 in C++14 mode, but not in C++11 mode.
The compiler may spill out a warning for \code{hst\_\-dev} 
if \code{func} takes its argument by value.
If the function \code{dev} is added,
the code does not compile any more under C++14, but compiles under C++17 with \nvcc 12.

Due to this quite erratic behaviour,
we advise against this solution.

\begin{figure}
\begin{tlisting}{{\color{red}Bad Solution}: Experimental relaxed \constexpr for lambdas (C++14)}{lst_solution_exp_constexpr_la_14}
#ifndef __CUDACC_RELAXED_CONSTEXPR__
#error Must be compiled with: \
  "--expt-relaxed-constexpr"
#endif

template< typename Lambda >
constexpr void func( Lambda && la ) {}

__host__ __device__ void hst_dev() {
  auto la = [](){};
  func( la );
}

//__device__ void dev() {
//  auto la = [](){};
//  func( la );
//}
\end{tlisting}
\Description{}
\squeezeup
\end{figure}

\subsubsection{Problem with extended lambdas}

Listing~\ref{lst_problem_lambda}, a modified version of~\cite{problem_lambda},
illustrates an issue with extended lambdas.

In the listing,
 a \device lambda is defined which captures the variable \code{h} by value.
Thus, the copy constructor of \code{H} needs to be called,
and gets instantiated.
Thus, \nvcc 12.1 gives the warning:
\compilermsg{%
call\-ing a \_\_host\_\_ function ("H::H(const H \&)") 
from a \_\_host\_\_ \_\_device\_\_ function
("main\-::\-[lam\-bda() (instance 1)\-]\-::\-[lamb\-da() (instance 1)]")
is not allowed}.
The compiler complains about the copy constructor of \code{H},
which is not marked with the \device annotation.\footnote{%
It is not clear yet to us, whether the copy constructor of \ccode{H}
is called in device code at all,
in which case the example would be ill-formed.
Adding suitable \ccode{printf} statements to the copy constructor of \ccode{H}
indicates that only the \chost copy constructor is ever called.
But, extended lambdas and kernel calls can involve some memcpying of data,
bypassing the copy constructor. This is against the C++ standard
but abides by the Cuda standard~\cite{cuda_doc_copying}.
}

\newsavebox{\LstBoxA}
\begin{lrbox}{\LstBoxA}
\begin{slisting}{}
template< bool b >
struct Foo {
  Foo() = default;
  Foo( const Foo & ) requires(  b ) {}
  Foo( const Foo & ) requires( !b ) {}
};

template< typename Lambda >
void wrapper( Lambda ) {}

int main() {
  Foo< true > foo;
  auto la = [foo](){};
  wrapper( la );
}
\end{slisting}
\end{lrbox}
Unfortunately, due to (potential) compiler bugs of \nvcc
we could not find a nice solution for this problem yet.\footnote{%
Since the problem is with the copy constructor, we need to use a \crequires clause.
Yet, \nvcc seems to have problems with that.
In particular, the following code does not compile under \nvcc
(but compiles under \gcc and \clang).

\noindent
\usebox{\LstBoxA}
}

\begin{figure}
\begin{tlisting}{Problem $\boldsymbol{\lambda}$}{lst_problem_lambda}
struct H {
  H() = default;      
  __host__ H( const H & ) {}
  __host__ __device__ void call() const{}
};

template< typename Lambda > __global__ 
void kernel( Lambda la ){ la(); }

int main() {
  H h;
  auto la = [h] __device__ () {
    h.call();
  };
  kernel<<< 1, 1 >>>( la );
  return cudaDeviceSynchronize();
}
\end{tlisting}
\Description{}
\squeezeup
\end{figure}


\subsection[SFINAE 1*]{SFINAE 1$^{\ast}$}
We now come to the more ugly solutions (involving macros).
Since the very resulting code is rather long,
we present it in steps.

\label{sol_sfinae_1s}
\subsubsection{Context}
We want to make sure at \emph{compile time} that a function which is either usable on the host or device side,
is only called on the host or device side.

\subsubsection{Solution}
We define three versions of the same function -- a \host, a \device, and a \host \device function --
and make it such that always only one (the one that we need) is well formed, and thus can be called.
In particular, the compiler cannot instantiate a wrong function and we do not get compiler warnings.
If we would try to do a stray function call we get a compilation error.
There is one gotcha: this leads to code duplication.

\paragraph{SFINAE and concepts}
Up to C++17 we have to use SFINAE\footnote{%
SFINAE: Substitution failure is not an error, which means
that an invalid substitution of template parameters does not yield a compilation error.
For a more thorough explanation see~\cite{cppref_sfinae,wiki_sfinae}.
}
for this task,
starting from C++20 we can use \emph{concepts}.
Since SFINAE clauses are usually very ugly
(and incomprehensible for people not familiar with it)
we wrap them in the helper macro \REQUIRES,
see Listing~\ref{lst_requires_11}.
An example how C++20 concepts are used is given in Listing~\ref{lst_requires_20}.

\paragraph{HDC$\,^\text{\textregistered}$}
We use a scoped enum \HDC (short for: \emph{Host Device Compatibility})
to determine which function needs to be called,
see Listing~\ref{lst_hst_dev_tags}.%
\footnote{%
We note again that, only due to restricted space we use very short names here.
In production code, some more descriptive names should be used.}
\begin{figure}
\begin{tlisting}{Host Device Compatibility}{lst_hst_dev_tags}
enum class HDC { Hst, Dev, HstDev };
\end{tlisting}
\Description{}
\squeezeup
\end{figure}

The first solution works the following:
The caller passes a compile time constant to our function
(either \code{HDC::Hst}, \code{HDC::Dev} or \code{HDC::HstDev}),
and the compiler then only instantiates a
\host, a \device, or a \host \device version of that function,
see Listing~\ref{lst_hd_macro_1s} for a C++14 solution.

\paragraph{Helper trait \hdc}
Since passing a hard coded compile time constant is of not much use,
we use a helper trait \code{hdc<>} which determines for a given type,
which function shall be instantiated.
\code{hdc< T >} inspects for the given type \code{T},
whether a member variable of name \code{T::hdc} is present.
If so, then \code{hdc< T >} returns its value.
Otherwise it returns \code{HDC::Hst},
see Listing~\ref{lst_hdc}.

\begin{figure}
\begin{tlisting}{host device macro 1$^\ast$}{lst_hd_macro_1s}
template<HDC x, REQUIRES(x==HDC::Hst)>
__host__ void f1s() {/*body*/}
template<HDC x, REQUIRES(x==HDC::Dev)>
__device__ void f1s() {/*body*/}
template<HDC x, REQUIRES(x==HDC::HstDev)>
__host__ __device__ void f1s() {/*body*/}

// Usage:
void g1a() {
  f1s< HDC::Hst >();
}

// Usage: with helper macro:
template< typename T >
void g1b() {
    f1s< hdc<T> >();
}
\end{tlisting}
\Description{}
\squeezeup
\end{figure}

\subsubsection{Assessment}

\newsavebox{\LstBoxB}
\begin{lrbox}{\LstBoxB}
\begin{slisting}{}
template< typename T >
struct C {
  __host__ C( C & ) requires
    ( hdc<T> == HDC::Hst ) { /*body*/ }
  __device__ C( const C & ) requires
    ( hdc<T> == HDC::Dev ) { /*body*/ }
  __host__ __device__ C( C & ) requires
    ( hdc<T> == HDC::HstDev ) { /*body*/ }
};
\end{slisting}
\end{lrbox}

\begin{itemize}
\item[$\circ$] This solution only works for template functions or functions of template classes.
This has some severe implications:
\begin{itemize}
\item[$+$] More possibilities for compiler optimizations.
\item[$-$] Source is physically coupled tighter, potentially leading to longer compilation times.
\item[$-$] The definition of the function must go into a header file.
\item[$-$] More code in header files and thus compilation times may increase.
\end{itemize}
\item[$\circ$] The \REQUIRES version is only applicable to template functions,
respectively it cannot be used when the function in question cannot be made into a template.
In particular, it is not\footnote{%
To overcome this problem, the standard approach up to C++17 
is to inherit from a templated class which has non-templated copy constructors.
This is so messy, that we refrain of giving a code example.
The much easier solution is to use C++20 -- or to wait until one can use C++20.
}
applicable to some special member functions of template functions,
e.g.\ the copy constructor.

The \requires version should work also in this case.\footnote{%
Here is a draft of a C++20 solution for a copy constructor.
It should work -- But:
\nvcc 12.2 seems to have bug and was not be able to compile this example.
In particular, I could not test whether this solution works actually.
\usebox{\LstBoxB}
}

\item[$-$] Code bloat and thus increased compilation times.
\item[$-$] Code duplication
\end{itemize}


\subsection[SFINAE 1+]{SFINAE 1$^{\dagger}$}
\label{sol_sfinae_1d}
\subsubsection{Context}
We have a class with template functions and
want to relieve the user to manually pass a template argument.

\subsubsection{Solution}
We use default template arguments
and define some constant at some suitable place,
see Listing~\ref{lst_hd_macro_1d}%
\footnote{%
It is necessary to introduce the name \ccode{c} in the Listing,
since using \ccode{hdc\_} in the \cREQUIRES clause would lead to a hard compilation error.
}
for a C++20 version.\XX{Is `x` for C++20 version necessary?}

\begin{figure}
\begin{tlisting}{host device 1$^\dagger$ (C++20)}{lst_hd_macro_1d}
template< HDC hdc_ >
struct S1d {
  template< HDC x = hdc_ >
    requires( x == HDC::Hst ) >
    __host__ void call(){}
  template< HDC x = hdc_ >
    requires( x == HDC::Dev ) >
    __device__ void call(){}
  template< HDC x = hdc_ >
    requires( x == HDC::HstDev ) >
    __host__ __device__ void call(){}
};

// Usage:
__device__ void g1d() {
  S1d< HDC::Dev > x;
  x.call();
}
\end{tlisting}
\Description{}
\squeezeup
\end{figure}

\subsubsection{Assessment}
Apart from the assessment in the solution before, we have:
\begin{itemize}
\item[$-$] Only works when we can define the variable \hdc in a sensible way,
this means effectively in a class.\footnote{%
In theory, it would also work in namespaces, or even at global scope, where also can easily define constants.
But namespaces (let alone the global scope) cannot be templatized.}
\end{itemize}


\subsection[SFINAE 1* with macro]{SFINAE 1$^\ast$ with macro}
\label{sol_sfinae_1_macro}

\subsubsection{Context}
We like solution SFINAE 1$^\ast$ but do not like that we have to duplicate code manually three times.
Furthermore, we have no objections against macros.

\subsubsection{Solution}
We use macros to automatically generate the boilerplate code,
see Listing~\ref{lst_hd_macro_1}.\footnote{%
Since macros are sensitive to commas, and we cannot use \VAARGS since our macro has multiple arguments,
we use the helper macro \DEPAREN.
\DEPAREN removes one pair of parentheses from its argument, if there is a pair,
see Listing~\ref{lst_deparen}.
This way, the user can decide whether she wants to parenthesize the passed macro arguments or not.
In simple cases, parentheses may not be necessary, and omitting them may increase readability.}

\begin{figure}
\begin{tlisting}{host device macro 1}{lst_hd_macro_1}
#define host_device_macro1( body )   \
  template< HDC hdc_ = hdc,          \
    REQUIRES( hdc_ == HDC::Hst ) >   \
  __host__ DEPAREN(body)             \
  template< HDC hdc_ = hdc,          \
    REQUIRES( hdc_ == HDC::Dev ) >   \
  __device__ DEPAREN(body)           \
  template< HDC hdc_ = hdc,          \
    REQUIRES( hdc_ == HDC::HstDev ) > \
  __host__ __device__ DEPAREN(body)

// Function definition:
template< HDC x >
struct S1 {
  static constexpr HDC hdc = x;
  host_device_macro1(
    (void call() {} )
  )
};

// Usage:
void g1() {
  S1< HDC::Hst > s;
  s.call();
}
\end{tlisting}
\Description{}
\squeezeup
\end{figure}

\subsubsection{Assessment}
Apart from the points in the solutions before, we have:
\begin{itemize}
\item[$+$] Duplicated code is automatically generated.
\item[$-$] Since the body of the function is in a macro, debugging is hard.\footnote{%
Actually, we do not know of any compiler which allows debugging a function defined in a macro.}
\end{itemize}


\subsection{SFINAE 3 with macro}
\label{sol_sfinae_3_macro}
\subsubsection{Context}
We like solution SFINAE 1$^*$ with macros,
but also want to use it with free standing functions.
Furthermore, we have no objections against macro abominations.

\subsubsection{Solution}
Instead of instantiating a class,
we deduce the execution space of a function from its arguments,
see Listing~\ref{lst_hd_macro_3}.
Again we use macros to generate all boilerplate code.

\begin{figure}
\begin{tlisting}{host device macro 3}{lst_hd_macro_3}
#define host_device_macro3(          \
    templateargs, hdc_, body )       \
  template< DEPAREN(templateargs),   \
    HDC hdc = DEPAREN(hdc_),         \
    REQUIRES( hdc == HDC::Hst) >     \
  __host__ DEPAREN(body)             \
  template< DEPAREN(templateargs),   \
    HDC hdc = DEPAREN(hdc_),         \
    REQUIRES( hdc == HDC::Dev) >     \
  __device__ DEPAREN(body)           \
  template< DEPAREN(templateargs),   \
    HDC hdc = DEPAREN(hdc_),         \
    REQUIRES( hdc == HDC::HstDev ) > \
  __host__ __device__ DEPAREN(body)

// Usage:
host_device_macro3( (typename T),
                    (hdc< T >),
                    (void f3( T t ) {}) )  

struct HD {
  static constexpr HDC hdc = HDC::HstDev;
};

void g3() {
  f3( 0 );     // calls host version
  f3( HD{} );  // calls device version
}
\end{tlisting}
\Description{}
\squeezeup
\end{figure}

\subsubsection{Assessment}
Apart from the points in the solutions before, we have:
\begin{itemize}
\item[$+$] Very easy to use for the user
\item[$\circ$] Reasonably easy to use for the implementer.
\end{itemize}

\subsubsection{Known usages}
\begin{itemize}
\item Dimetor (Closed source code)
\end{itemize}

\begin{figure}
\begin{tlisting}{Solution: SFINAE 3}{lst_solution_sfinae_3}
struct D {
  static constexpr HDC hdc = HDC::Dev;
  __device__ void call() {}
};

struct H {
  __host__ void call() {}
};

host_device_macro3(
  (typename T),
  (hdc< T >),
  (void func( T t ) { t.call(); }) )
  
__global__ void kernel() {
  // func( H{} );  // does not compile
  func( D{} );
}

int main() {
  func( H{} );
  // func( D{} );  // does not compile
}
\end{tlisting}
\Description{}
\squeezeup
\end{figure}


\subsection{SFINAE with pragma}
\label{sol_sfinae_pragma}

\subsubsection{Context}
We like solution SFINAE 3, but also want to be able to debug the code,
since non debuggable code is a no go in most situations.
Furthermore, we have no objections against even larger macro abominations.

\subsubsection{Solution}
First we use the SFINAE approach to guide the compiler to a suitable annotated wrapper function.
The wrapper function is generated using a macro, as before.
The wrapper function itself forwards the call to a \host \device function.

This alone does not solve the problem, since we are now calling again unconditionally a
\host \device function. 
Thus, the compiler will spit out the same warning we had at the beginning.
But due to our indirection, we know that we cannot make a stray function call,
and therefore, we can safely silence those warnings with the pragmas
from Solution~\ref{sol_pragma}.
\footnote{%
nvcc implicitly considers \cstdmove and \cstdforward to have \chost \cdevice annotations.
Therefore, the code in Listing~\ref{lst_solution_sfinae_pragma}
is valid.~\cite{cuda_doc_move_forward}
}

Furthermore, from our tests described in Section~\ref{sec_function_exectuion_space_specifiers}-``\nameref{sec_function_exectuion_space_specifiers}''
we can infer that a stray call leads to a compilation error.
Indeed, for a stray call we have either a call of ``host $\rightarrow$ device'',
or ``device $\rightarrow$ host'',
which in both cases leads to an error.

See Listing~\ref{lst_solution_sfinae_pragma} for the (C++20) solution with the expanded macro,
and Listing~\ref{lst_hd_pragma_macro_4} for the macro itself
and how to use it. One can see, that this solution again needs boilerplate code --
One has to write the code to forward the arguments to the actual function 
(here: \code{func\_imp}) by hand.
Also, one should not forget to add the pragma.
If one forgets it, compiler warnings will occur, but one is still safe from UB.

\begin{figure}
\begin{tlisting}{Solution: SFINAE with pragma}{lst_solution_sfinae_pragma}
struct H {
  void call() {}
};

struct D {
  static constexpr HDC hdc = HDC::Dev;
  __device__ void call() {}
};

#pragma nv_exec_check_disable
template< typename T >
__host__ __device__
void func_impl( T && t ) {
  t.call();
}

/* macro generated: start */
template< typename T, HDC hdc = hdc<T> >
requires( hdc == HDC::Hst )
__host__ void func( T && t ) {
  return func_impl( std::forward<T>(t) );
}
template< typename T, HDC hdc = hdc<T> >
requires( hdc == HDC::Dev )
__device__ void func( T && t ) {
  return func_impl( std::forward<T>(t) );
}
template< typename T, HDC hdc = hdc<T> >
requires( hdc == HDC::HstDev )
__host__ __device__ void func( T && t ) {
  return func_impl( std::forward<T>(t) );
}
/* macro generated: end */

// Usage:
__global__ void kernel() {
//func( H{} );  // error
  func( D{} );
}

int main() {
  func( H{} );
//func( D{} );  // error
}
\end{tlisting}
\Description{}
\squeezeup
\end{figure}

\begin{figure}
\begin{tlisting}{host device pragma macro~4}{lst_hd_pragma_macro_4}
#define host_device_pragma_macro4(   \
    TemplArg, hd_, sig, wrapper )    \
  template< DEPAREN(TemplArg),       \
            HDC hdc = DEPAREN(hd_) > \
  requires( hdc == HDC::Hst )        \
  __host__ DEPAREN(sig) {            \
    DEPAREN(wrapper) }               \
  template< DEPAREN(TemplArg),       \
            HDC hdc = DEPAREN(hd_) > \
  requires( hdc == HDC::Dev )        \
  __device__ DEPAREN(sig) {          \
    DEPAREN(wrapper) }               \
  template< DEPAREN(TemplArg),       \
            HDC hdc = DEPAREN(hd_) > \
  requires( hdc == HDC::HstDev )     \
  __host__ __device__ DEPAREN(sig) { \
    DEPAREN(wrapper) }

// Usage:
host_device_pragma_macro4(
  (typename T ),
  (hdc< T >),
  (void func( T && t )),
  (return func_impl(
      std::forward<T>(t) ); )
)
\end{tlisting}
\Description{}
\squeezeup
\end{figure}

\subsubsection{Assessment}
\begin{itemize}
\item[$+$] Code is debuggable.
\item[$+$] Easy to use for the user.
\item[$\circ$] Library implementer again has to write some boilerplate code.
\item[$-$] Not straight forward to write and understand, and thus hard to maintain in the long term.
\item[$-$] One indirection for each function call in debug mode via \codestl{std::forward}.
For implications of this, see~\cite[Item 30]{Meyers_eff_modern}.
In Release mode this indirection most likely is optimized out.
\end{itemize}

\subsubsection{Known usages}
No known usages.


\setcounter{figure}{100}

\appendix
\section{Patterns which need language changes}
We now present patterns which would allow to solve the problem elegantly,
but need changes to the Cuda language.
The proposals are currently under review at Nvidia.

Although the solutions are split in two,
the proposed language changes are mostly mutually independent from,
and compatible to each other.

\subsection{Proposal: Conditional \host \device}
This solution needs minimal language changes
and does not introduced any breaking change in the language.
The solution is similar to the SFINAE Solution~\ref{sol_sfinae_3_macro}.

We propose that \device (and \host ) annotations except a boolean parameter. 
If a parameter is given , and it is false,
then the function is not compiled for \host (or \host).\footnote{%
A similar approach is taken in C++ with the \cnoexcept,
or the \cexplicit (C++20) keyword,
which also takes an optional boolean argument.}
See Listing~\ref{lst_proposal_1} how to use it
to solve our Problem T from the beginning.\footnote{%
In the listing we use the \chdc helper function.
If this proposal is considered to be accepted,
then one should probably rethink the API of the \chdc helper function.}

\begin{figure}
\begin{tlisting}{Proposal 1}{lst_proposal_1}
struct H {
  __host__ void call() {}
};

struct D {
  HDC hdc = HDC::Dev;
  __device__ void call() {}
};

template< typename T >
__host__( hdc<T>==Hst::Hst )
__device__( hdc<T>==Hst::Dev )
void wrap() {
  T{}.call();
}

int main() {
//wrap< D >(); // error
  wrap< H >(); // OK
}
\end{tlisting}
\Description{}
\squeezeup
\end{figure}

\subsubsection{Assessment}
\begin{itemize}
\item[$+$] Needs minimal language changes.
\item[$+$] Easy to use. 
\item[$\circ$] Only applicable to template functions or member functions of template classes.
\end{itemize}

\subsection{Proposal: Propagation of \host \device}
This solution solves our problem%
,
but also needs more language changes.
It also imposes more restrictions to the programming style.
In particular, a lot of functions must go to a header file.

We discuss each (potentially) necessary language change on its own.

\subsubsection{Allowing \host \device decorations for classes and namespaces}
If a class (or name\-space) is decorated with 
\device (or \host \device),
then all its undecorated member functions
become automatically \device (or \host \device) member functions.

If a function cannot be implicitly decorated, then the compilation fails.

For example, the two classes \code{S1} and \code{S2} in Listing~\ref{lst_hst_dev_class} are equivalent.

\begin{figure}
\begin{tlisting}{\host \device decorations for classes and namespaces}{lst_hst_dev_class}
__device__ class S1 {
  void call();
  __host__ void init();
};
class S2 {
  __device__ void call();
  __host__ void init();
};
\end{tlisting}
\Description{}
\squeezeup
\end{figure}

\subsubsection{Including \host \device in the signature of the function}
We propose that the function execution space specifiers are part of the signature.
Them, one dan get rid of using the
\CUDAARCH macro in cases where the host and device part need different code,
see Listing~\ref{lst_function_signature}.

The function \code{g} calls \code{f\coderemark{/*(1)*/}} when \code{g} is compiled for \host code,
and calls \code{f\coderemark{/*(2)*/}} when \code{g} is compiled for \device code.\footnote{%
For the \nvcc compiler the \chost, \cdevice annotations are not part of the function signature;
whereas clang considers them to be part of the signature.}

\begin{figure}
\begin{tlisting}{\host \device as part of the function signature}{lst_function_signature}
__host__ void f/*(1)*/() {}
__device__ void f/*(2)*/() {}

__host__ __device__ void g() {
  func();
}
\end{tlisting}
\Description{}
\squeezeup
\end{figure}

\subsubsection{Propagation of \host \device}
If a \device or \cudaglobal, (or \host \device) function calls
an undecorated function
(or an undecorated member function in an undecorated class),
it becomes automatically a \device (or \host \device) function.
If a \device, or \cudaglobal, (or \host \device) function calls
an undecorated templated (member) function,
then this function is instantiated as a \device (or \host \device) function.

This implies,
$(1)$ that the \host and \device annotations need to be part of the function signature, 
(otherwise the compiler cannot decide which function to call), and
$(2)$ that the definition of the function must be present to compiler
whenever the function is invoked
(otherwise, the compiler does not know which functions shall be instantiated and a linker error occurs).

$(2)$ implies that most functions must go to header files.
To mitigate this shortcoming,
one could allow \emph{explicit \host and \device instantiations}.\footnote{%
The explicit instantiation would be similar to explicit C++ template function instantiations.}

\subsubsection{Forbid stray function calls}
We propose that the compiler shall handle stray function calls consistently and strictly.
In particular, every call of a \host to a \device function
(and of a \device to a \host function)
shall be a compiler error.
Every call of a \host \device to a \host or \device function shall be at least a compiler warning.%
%

\subsubsection{Solution}
See Listing~\ref{lst_proposal_2} how to use this proposal to solve our Problem T from the beginning.

\begin{figure}
\begin{tlisting}{Proposal 2}{lst_proposal_2}
struct S { void call() {} };
__host__ struct H { void call() {} };
__device__ struct D {
  HDC hdc = HDC::Dev;
  void call() {}
};

template< typename T >
void wrap() {
  return T{}.call();
}

__global__ void kernel() {
  wrap< S >(); // OK
  wrap< H >(); // error
}

int main() {
//wrap< D >(); // error
  wrap< H >(); // OK
}
\end{tlisting}
\Description{}
\squeezeup
\end{figure}

\subsubsection{Assessment}
\begin{itemize}
\item[$+$] Adding the \host \device decorators to the function signature simplifies learning Cuda.
Often novices of Cuda/C++ think that the \host \device annotations are part of the function signature.
\item[$+$] Allowing decorations on classes allows to port code more easily to Cuda.
\item[$+$] Needs no further work from the user.
\item[$+$] Is applicable to third party code.
\item[$\circ$] Functions then behave like template functions with all implications.
\item[$-$] Code must be put into header files, like for template functions and classes.
\end{itemize}

\section{Helper functions and macros}

\subsubsection{Host Device compatibility (\hdc)}
\code{hdc< T >} inspects for the given type \code{T},
whether a member variable of name \code{T::hdc} is present.
If so, then \code{hdc< T >} returns its value.
Otherwise it returns \code{HDC::Hst},
see Listing~\ref{lst_hdc}.

Depending on the use case, 
one may want to alter the code such that \hdc returns \code{HDC::HstDev}
for fundamental types like \codekeyword{int},... and trivial C-style structs.
To this end,
the type traits \codestl{std::is\_\-fun\-da\-men\-tal} and \codestl{std::is\_triv\-ial} can be used.

\begin{figure}
\begin{tlisting}{Host Device compatibility: \code{hdc}}{lst_hdc}
#include <type_traits>

HAS_MAKE( hdc )

template< typename T >
struct hdc_impl {
  struct h {
    static constexpr HDC hdc = HD::Hst;
  };
  static constexpr HDC hdc =
    std::conditional_t< has_hdc<T>,
                        T, h >::hdc;
};

template< typename T > static constexpr
HD hdc = hdc_impl< T >::hdc;

// Usage example
struct S {};
struct D {
  static constexpr HDC hdc = HDC::Dev;
};

static_assert( hdc<S> == HDC::Hst );
static_assert( hdc<D> == HDC::Dev );
\end{tlisting}
\Description{}
\squeezeup
\end{figure}

\subsubsection{\REQUIRES}
Since Cuda lacks behind the C++ standard usually for years
and/or one cannot always use a compiler which supports the latest C++ standard,
we will make use of a~\REQUIRES macro,
which simplifies writing SFINAE clauses enormously.
The macro expands to a form usage of the \stdenableif pattern~\cite{cppref_enable_if},
see Listing~\ref{lst_requires_11}.\footnote{%
The macro uses~\cVAARGS in order to be able to handle arguments which include a ``\ccode{,}''.}
For comparison, a C++20 version is given in Listing~\ref{lst_requires_20}.

The listings also give an example of a function \code{inc} which increments the input by 1
and which is only applicable to integral types.

\begin{figure}
\begin{tlisting}{\REQUIRES (C++11)}{lst_requires_11}
#include <type_traits>

#define REQUIRES( ... )    \
  typename std::enable_if< \
    __VA_ARGS__, bool      \
  >::type = false

// usage example
template< typename T,
    REQUIRES( std::is_integral_v<T> ) >
T inc( T i ) { return i + 1; }
\end{tlisting}
\Description{}
\squeezeup
\end{figure}

\begin{figure}
\begin{tlisting}{\requires (C++20)}{lst_requires_20}
#include <concepts>

template< typename T >
requires std::integral< int >
T inc( T i ) { return i + 1; }
\end{tlisting}
\Description{}
\squeezeup
\end{figure}

\subsubsection{\DEPAREN}
This is a helper macro to remove parentheses from macro arguments.

\begin{figure}
\begin{tlisting}{\DEPAREN}{lst_deparen}
#define DEPAREN(X) ESC(ISH X)
#define ISH(...)   ISH __VA_ARGS__
#define ESC(...)   ESC_(__VA_ARGS__)
#define ESC_(...)  VAN ## __VA_ARGS__
#define VANISH

// usage example
DEPAREN( (int) ) a;  // gives: int a;
DEPAREN( int ) b;    // gives: int b;
\end{tlisting}
\Description{}
\squeezeup
\end{figure}

\subsubsection{\hasxxx type trait}
\hasxxx returns wheth\-er a class has a certain member function or variable of a certain name.
In production code one better renames it to \code{has\_member\_xxx}.
We give three very different versions.

A C++14 version is given in Listing in~\ref{lst_has_make_14}~\cite{has_macro_14}.
It is quite verbose 
and for each \emph{name} one has to generate a struct using the
\HASMAKE macro.
There is furthermore a major difference to the other two versions.
This code checks whether a certain \emph{name} is defined in the class in question,
which can also be typedefs, unscoped enum values etc., and also works for private members.

The C++17 version has the advantage,
that one does not need to generate a type trait for each name.
For compatibility reasons to our code listing,
we nevertheless define such a helper trait,
see Listing~\ref{lst_has_make_17}~\cite{has_macro_17}.

\begin{figure}
\begin{tlisting}{\has (C++17)}{lst_has_make_17}
#include <type_traits>

template< typename T, typename Lambda >
constexpr auto has_( Lambda && la )
 -> decltype(la(std::declval<T>()), true)
{ return true; }

template< typename >
constexpr auto has_( ... ) -> bool
{ return false; }

#define has( T, EXPR ) \
 has_< T >( \
 [](auto && obj)->decltype(obj.EXPR){} )

// For compatibility to C++14/20 version
#define HAS_MAKE( name ) \
 template<typename T> static constexpr \
 bool has_ ## name = has( T, name );

// usage example
struct X {
  int foo;
};
static_assert( has(X, foo) );

HAS_MAKE( foo )
static_assert(  has_foo<X> );
static_assert( !has_foo<int> );
\end{tlisting}
\Description{}
\squeezeup
\end{figure}

A C++20 version is given in Listing in~\ref{lst_has_make_20}.
Similar to the C++14 version, for each name one has to generate a trait.
Apart from that, it is by far the easiest version.

\begin{figure}
\begin{tlisting}{\hasxxx (C++20)}{lst_has_make_20}
#define HAS_MAKE( name )  \
  template< typename T >  \
  concept has_ ## name =  \
  requires( T t ) { &T::name; };

HAS_MAKE( foo );
\end{tlisting}
\Description{}
\squeezeup
\end{figure}

\begin{figure*}
\begin{tlisting}{\hasxxx (C++14)}{lst_has_make_14}
#define HAS_MAKE( name )                                                    \
template< typename T > struct has_ ## name ## _c {                          \
  struct NameCollission { char name; };                                     \
  struct Both : T, NameCollission {};                                       \
  template< typename U, U > struct Check;                                   \
  template< typename U > static std::false_type test(                       \
      Check<char NameCollission::*, &U:: name>* );                          \
  template< typename U > static std::true_type test( ... );                 \
  static constexpr bool value = decltype( test<Both>(nullptr) )::value;     \
};                                                                          \
                                                                            \
template< typename T, REQUIRES( std::is_class<T>::value ) >                 \
constexpr int has_ ## name ## _f() { return has_ ## name ## _c<T>::value; } \
                                                                            \
template< typename T, REQUIRES( !std::is_class<T>::value ) >                \
constexpr int has_ ## name ## _f() { return false; }                        \
                                                                            \
template< typename T > constexpr bool has_ ## name = has_ ## name ## _f< T >();

// Usage:
HAS_MAKE( foo )

struct X {
    int foo();
};

static_assert( has_foo<X> );
static_assert( !has_foo<int> );
\end{tlisting}
\Description{}
\squeezeup
\end{figure*}

%
%
%

\section{Additional material}
Information about restrictions device code
can be found in the \emph{Cuda Documentation}~\cite[Section 14]{cuda_doc_14}.
Meyers (old, but still relevant) book \emph{Effective Modern C++} treats
C++14 language constructs~\cite{Meyers_eff_modern}.
Extensive documentation about C++ (including older versions) is given at
\emph{cppreference.com}~\cite{cppreference}.
The \emph{current working draft of the C++ standard} (and links to older drafts) 
are found at~\cite{cppdraft}.
Compilers, and different versions of it, can be tested using \emph{godbolt}~\cite{godbolt}.


\newcommand{\doi}[1]            {\href{https://doi.org/#1}{doi:~#1}}
\newcommand{\arxiv}[1]          {\href{https://arxiv.org/abs/#1}{arXiv:~#1}}
\newcommand{\cudadoc}[1]        {docs.n\-vidia.com/\-cu\-da/\-cu\-da-c-pro\-gram\-ming-gui\-de/\-in\-dex.html\#\-#1}
\newcommand{\stackoverflow}[1]  {stack\-over\-flow.com/\-#1}
\newcommand{\cppref}[1]         {cpp\-ref\-er\-ence.com/\-#1}
\newcommand{\wiki}[1]           {en.\-wiki\-pe\-dia.org/\-wiki/\-#1}
\newcommand{\gitlab}[1]         {git\-lab.com/\-#1}
\newcommand{\github}[1]         {git\-hub.com/\-#1}

\end{document}